\documentclass[aps,prl,twocolumn,superscriptaddress,preprintnumbers,%
               showpacs,nofootinbib]{revtex4}
\newcommand{\PRE}[1]{}       

\usepackage{bm}
\usepackage{epsfig}

\newcommand{\postscript}[2]{\setlength{\epsfxsize}{#2\hsize}
   \centerline{\epsfbox{#1}}}

\def\to{\rightarrow}

\def\bi{\begin{itemize}}
\def\ei{\end{itemize}}
\def\te{\tilde e}

\def\tu{\tilde u}

\def\tb{\tilde b}

\def\tst{\tilde t}
\def\ttau{\tilde \tau}

\def\tg{\tilde g}
\def\tnu{\tilde\nu}
\def\tell{\tilde\ell}
\def\tq{\tilde q}
\def\tw{\widetilde W}
\def\tz{\widetilde Z}
\def\CM{{\cal{M}}}
\def\alt{\stackrel{<}{\sim}}
\def\agt{\stackrel{>}{\sim}}
\def\be{\begin{equation}}  
\def\ee{\end{equation}}  
\def\bea{\begin{eqnarray}}  
\def\eea{\end{eqnarray}}

\newcommand\plb[3]{{\it Phys\ Lett.\ }{\bf B #1} (#2) #3}
\newcommand\jhep[3]{{\it J.\ High Energy Phys.\ }{\bf #1} (#2) #3}
\newcommand\npb[3]{{\it Nucl.\ Phys.\ }{\bf B #1} (#2) #3}

\newcommand\prD[3]{{\it Phys.\ Rev.\ }{\bf D #1} (#2) #3}
\begin{document}

\preprint{FTPI-MINN-12/22}
\preprint{UMN-TH-3109/12}
\preprint{UH-511-1195-12}

\title{
\PRE{\vspace*{1.5in}}
Radiative natural SUSY with a 125 GeV Higgs boson
\PRE{\vspace*{0.3in}}
}

\author{Howard Baer}
\affiliation{Dept. of Physics and Astronomy,
University of Oklahoma, Norman, OK, 73019, USA
\PRE{\vspace*{.1in}}
}
\author{Vernon Barger, Peisi Huang}
\affiliation{Dept. of Physics,
University of Wisconsin, Madison, WI 53706, USA
\PRE{\vspace*{.1in}}
}
\author{Azar Mustafayev}
\affiliation{W. I. Fine Institute for Theoretical Physics, 
University of Minnesota, Minneapolis, MN 55455, USA
\PRE{\vspace*{.1in}}
}
\author{Xerxes Tata}
\affiliation{Dept. of Physics and Astronomy,
University of Hawaii, Honolulu, HI 96822, USA
\PRE{\vspace*{.1in}}
}


\begin{abstract}
\PRE{\vspace*{.1in}} It has been argued that requiring low electroweak
fine-tuning (EWFT) along with a (partial) decoupling solution to the
SUSY flavor and $CP$ problems leads to a sparticle mass spectra
characterized by light Higgsinos at 100-300 GeV, sub-TeV third
generation scalars, gluinos at a few TeV and multi-TeV first/second
generation scalars (natural SUSY).  We show that by starting with
multi-TeV first/second and third generation scalars and trilinear soft
breaking terms, the natural SUSY spectrum can be generated radiatively
via renormalization group running effects. Using the complete 1-loop
effective potential to calculate EWFT, significantly heavier third
generation squarks can be allowed even with low EWFT. The large negative
trilinear term and heavier top squarks allow for a light Higgs scalar in
the $\sim 125$ GeV regime.
\end{abstract}

\pacs{12.60.-i, 95.35.+d, 14.80.Ly, 11.30.Pb}

\maketitle

Over 11 fb$^{-1}$ of data has now been collected at the CERN LHC, and
evidence at the $5\sigma$ level has emerged for the existence of a
Higgs-like boson with mass $m_h\simeq 125$ GeV\cite{atlash,cmsh}.  While
the Standard Model (SM) allows for a Higgs scalar anywhere within the
range $\sim 115-800$ GeV\footnote{The lower end of this mass range comes
from previous Higgs searches at the LEP2 collider\cite{leph}, while the
upper value comes from the classic unitarity limits\cite{lqt}.}  the
minimal supersymmetric Standard Model (MSSM) requires that $m_h\alt 135$
GeV\cite{mssmhreview}.  That the Higgs boson mass value falls within the
narrow MSSM window may be regarded at least as supportive evidence for
the existence of weak scale supersymmetry\cite{wss}.  However, during
the same data taking run of LHC, no signal for SUSY has
emerged\cite{atlassusy,cmssusy,ichep}, leading to mass limits of $m_{\tg}>1.4$
TeV for $m_{\tq}\sim m_{\tg}$, and $m_{\tg}\agt 0.85$~TeV when
$m_{\tq}\gg m_{\tg}$ within the popular minimal supergravity (mSUGRA or
CMSSM) model\cite{msugreview}.  These strong new sparticle mass limits
from LHC push models such as mSUGRA into rather severe conflict with
electroweak fine-tuning (EWFT) calculations\cite{bbhmt}, leading many
physicists to consider alternative SUSY models which allow for much
lower EWFT\cite{hs,pap,sundrum,essig,hall,ns,bad,randall,thaler}.

The EWFT arising in SUSY models can be gleaned most easily from the Higgs portion of the scalar potential, 
which in the MSSM is given by
\be
V_{Higgs}=V_{tree}+\Delta V,
\ee
where the tree level portion is given by
\bea
V_{tree}=(m_{H_u}^2+\mu^2)|h_u^0|^2 +(m_{H_d}^2+\mu^2)|h_d^0|^2 \nonumber \\ 
-B\mu (h_u^0h_d^0+h.c.)+{1\over 8}(g^2+g^{\prime 2})
(|h_u^0|^2-|h_d^0|^2)^2
\eea
and the radiative corrections (in the effective potential approximation) by
\be
\Delta V=\sum_{i}\frac{(-1)^{2s_i}}{64\pi^2}Tr\left((\CM_i\CM_i^\dagger)^2
\left[\log {\CM_i\CM_i^\dagger\over Q^2}-{3\over 2}\right]\right) ,
\ee
where the sum over $i$ runs over all fields that couple to Higgs fields,
$\CM_i^2$ is the {\it Higgs field dependent} mass squared matrix
(defined as the second derivative of the tree level Lagrangian) of each
of these fields, and the trace is over the
internal as well as any spin indices.
Minimization of the scalar potential in the $h_u^0$ and $h_d^0$ directions allows one to compute the 
gauge boson masses in terms of the Higgs field vacuum expectation values $v_u$ and $v_d$, and 
leads to the well-known condition that
\be
\frac{m_Z^2}{2} = \frac{(m_{H_d}^2+\Sigma_d^d)-(m_{H_u}^2+\Sigma_u^u)\tan^2\beta}{(\tan^2\beta -1)}
-\mu^2 
\label{eq:mssmmu},
\ee where the $\Sigma_u^u$ and $\Sigma_d^d$ terms arise from derivatives
of $\Delta V$ evaluated at the potential minimum and
$\tan\beta\equiv\frac{v_u}{v_d}$.  At the one-loop level, $\Sigma_u^u$
contains 18 and $\Sigma_d^d$ contains 19 separate contributions from
various particles/sparticles\cite{bbhmt}. 
This minimization condition relates the
$Z$-boson mass scale to the soft SUSY breaking terms and the
superpotential higgsino mass $\mu$.

In order for the model to enjoy {\it electroweak
naturalness}\footnote{Our definition of electroweak naturalness derives
directly from the relation Eq.~\ref{eq:mssmmu}, which only involves SUSY
parameters at the electroweak scale.  Alternatively, one may apply
fine-tuning considerations to how likely it is to generate specific weak
scale parameter sets from high scale model parameters, or on how
sensitive $M_Z$ is to GUT scale parameters.  
The hyperbolic branch/focus point (HB/FP) region of the mSUGRA model is
not fine-tuned with respect to the $\mu$-parameter, but the presence of
heavy third generation scalars requires large cancellations between
$m_{H_u}^2$ and $\Sigma_u^u$ terms in Eq.~(\ref{eq:mssmmu}).} we adopt a fine-tuning
measure which requires that each of the 40 terms on the right-hand-side
(RHS) of Eq.~(\ref{eq:mssmmu}) should be of order $\sim m_Z^2/2$.
Labeling each term as $C_i$ (with $i=H_d,\ H_u,\ \mu ,\
\Sigma_d^d(\tst_1),\ \Sigma_u^u(\tst_1),\ etc.$), we may require
$C_{max}\equiv max|C_i|<\Lambda_{max}^2$, where $\Lambda_{max}\sim
100-300$ GeV, depending on how much EW fine-tuning one is willing to
tolerate. This measure of fine-tuning is similar to (but not exactly the
same as) Kitano-Nomura\cite{kn} but different from
Barbieri-Giudice\cite{bg} beyond the tree-level.
\footnote{Barbieri and Giudice\cite{bg} define a fine tuning measure
$\Delta_{BG}=max|(a_i/M_Z^2)\partial M_Z^2/\partial a_i|$ for
input parameters $a_i$. 
If we apply this to weak scale parameters $\mu^2$ or $m_{H_u}^2$ in 
Eq.~(\ref{eq:mssmmu}), 
our EWFT measure coincides with theirs at tree-level but differs when
radiative corrections embodied in the $\Sigma$ terms are included. 
In models defined at the high scale there are additional contributions to 
finetuning from corrections involving large logarithms that show up
in $\Delta_{\rm BG}$ applied to $m_{H_u}^2(M_{GUT})$.
Details will be presented in a future publication\cite{bbhmmt}.}
We will define the new fine-tuning parameter
$\Delta =C_{max}/(m_Z^2/2)$, where lower values of $\Delta$ correspond
to less fine-tuning, and {\it e.g.} $\Delta =20$ would correspond to
$\Delta^{-1}=5\%$ fine-tuning.

If we now require $\Lambda_{max}=200$ GeV ({\it i.e.} $\Delta \alt 10$
or $\Delta^{-1}>10\%$ fine-tuning), we find that $|\mu |<200$ GeV,
leading to a SUSY spectrum with light higgsinos in the range 100-200
GeV.  Since the terms in Eq.~(\ref{eq:mssmmu}) involving $m_{H_d}^2$ and
$\Sigma_d^d$ are suppressed by $\tan^2\beta$, $m_{H_d}^2$, and hence
$m_A^2$, can be large without violating our fine tuning criterion; in
this case, $m_A \alt \Lambda_{\rm max}\tan\beta$.
The largest of the radiative corrections in $\Sigma_u^u$ is expected to
come from top squarks: $\Sigma_u^u\sim \frac{3}{16\pi^2}f_t^2
F(m_{\tst_{1,2}}^2)$, where $F(m^2)=m^2(\log \frac{m^2}{Q^2}-1)$ and
$f_t$ is the top quark Yukawa coupling.  Requiring
$\Sigma_u^u\alt\Lambda_{max}^2=(200\ {\rm GeV})^2$ and assuming
$F(m^2)\sim m^2$ then seemingly implies a spectrum of light top squarks
$m_{\tst_{1,2}}\alt 1.5$ TeV and by $SU(2)$ symmetry, $m_{\tb_L}\alt
1.5$ TeV. Since the gluino loop contribution to the top squark mass goes
like $\delta m_{\tst_i}^2\sim \frac{2g_s^2}{3\pi^2}m_{\tg}^2\times log$,
where the $log\sim 1$, we also get a bound that $m_{\tg}\alt
3m_{\tst_i}\alt 4.5$~TeV. Thus, the sparticle mass spectra, here known
as {\it natural SUSY}\cite{kn,pap,sundrum,ns}, is characterized by (in
the case where $\Lambda_{max}=200$ GeV) \bi
\item higgsino-like charginos $\tw_1$ and neutralinos $\tz_{1,2}$ with mass $\alt 200$ GeV,
\item third generation squarks $m_{\tst_{1,2}},\ m_{\tb_1}\alt 1.5$ TeV,
\item $m_{\tg}\alt 3-4.5$ TeV, depending on $m_{\tst_1}$.  \ei Since
first/second generation Yukawa couplings are tiny, the first/second
generation squarks and sleptons enter $\Sigma_u^u$ with only tiny
contributions, so that their masses can be pushed into the multi-TeV
regime, offering at least a partial decoupling solution to the SUSY
flavor and $CP$ problems\cite{dine,am,nelson,esusy}. Thus, it is also
possible that \bi
\item $m_{\tq_{1,2}},\ m_{\tell_{1,2}}\sim 10-20$ TeV,
\ei
which is well beyond LHC search limits. 

Numerous recent papers have been published examining aspects of natural
SUSY. Regarding collider searches for natural SUSY, the light higgsinos
can be produced at LHC at appreciable rates, but their small mass gaps
$m_{\tw_1}-m_{\tz_1}\sim m_{\tz_2}-m_{\tz_1}\sim 10-20$ GeV lead to very
soft visible energy release which is hard to detect above SM background
at LHC\cite{hs}.  The light third generation squarks, gluinos and
heavier electroweak-inos
may not be accessible to LHC searches depending on their masses and
decay modes. A definitive test of natural SUSY may have to await
searches for the light higgsino-like charginos and neutralinos at an
International Linear $e^+e^-$ Collider (ILC), which in this case would
be a {\it higgsino factory}, in addition to a Higgs
factory\cite{tadas,hs,ns,blist}.

While the advantages of natural SUSY are clear (low EWFT, decoupling
solution to SUSY flavor and $CP$ problems), some apparent problems seem
to arise. First among these is that the sub-TeV spectrum of top squarks
feed into the calculation of $m_h$, usually leading to $m_h$ in the
115-120 GeV range, rather than $m_h\simeq 125$ GeV. Put more simply, a
value $m_h\sim 125$ GeV favors top squark masses in excess of 1
TeV\cite{h125}, while natural SUSY expects top squark masses below the
TeV scale. A separate issue is the apparent disparity
between the TeV third generation scale and the 10-20~TeV
first/second generation mass scale; we will illustrate that it is
possible to generate this radiatively. Several papers have appeared which
attempt to reconcile the large value of $m_h$ with naturalness by adding
extra singlet fields to the theory, which provide extra contributions to
$m_h$, thereby lifting it into its measured
range\cite{hall,nev,randall}.  This is what occurs in the
NMSSM\cite{nmssm}.  This solution may not be as appealing as it sounds
in that additional singlets can destabilize the gauge hierarchy via
tadpole effects\cite{bpr}, and may lead to cosmological problems via
domain walls\cite{domwall}.  In this paper, we reconcile a large value
of $m_h\sim 123-127$ GeV with low EWFT, and at the same time avoid at
least a gross disparity between the soft breaking matter scalar mass
scales, all the while avoiding the introduction of extra gauge singlets
or any other sort of exotic matter.

To begin with, we return to our measure of EWFT: $\Delta =C_{max}/(m_Z^2/2)$. We calculate the complete
1-loop effective potential contributions to the quantities $\Sigma_d^d$ and $\Sigma_u^u$ in Eq.~(\ref{eq:mssmmu}).
We include contributions from $W^\pm$, $Z$, $\tst_{1,2}$, $\tb_{1,2}$, $\ttau_{1,2}$, $\tw_{1,2}$, 
$\tz_{1,2,3,4}$, $t$, $b$ and $\tau$, $h$, $H$ and $H^\pm$. We adopt a scale choice
$Q^2=m_{\tst_1}m_{\tst_2}$ to minimize the largest of the logarithms. Typically, the largest
contributions to $\Sigma_u^u$ come from the top squarks, where we find 
\bea
\Sigma_u^u(\tst_{1,2} )&=&\frac{3}{16\pi^2}F(m_{\tst_{1,2}}^2)\times \nonumber \\
\left[ f_t^2-g_Z^2\right. &\mp &
\left.\frac{f_t^2 A_t^2-8g_Z^2(\frac{1}{4}-\frac{2}{3}x_W)\Delta_t}{m_{\tst_2}^2-m_{\tst_1}^2}
\right]
\label{eq:Siguu}
\eea
where $\Delta_t=(m_{\tst_L}^2-m_{\tst_R}^2)/2+m_Z^2\cos
2\beta(\frac{1}{4}-\frac{2}{3}x_W)$, $g_Z^2=(g^2+g^{\prime 2})/8$ and
$x_W\equiv \sin^2\theta_W$. This equation is somewhat more complicated
than the naive expression mentioned earlier, and contains contributions
from the $A_t$ parameter.  For the case of the $\tst_1$ contribution, as
$|A_t|$ gets large there is a suppression of $\Sigma_u^u(\tst_1)$ due to
a cancellation between terms in the square brackets of
Eq.~(\ref{eq:Siguu}).  For the $\tst_2$ contribution, the large
splitting between $m_{\tst_2}$ and $m_{\tst_1}$ yields a large
cancellation within $F(m_{\tst_2}^2)$
$\left(\log(m_{\tst_2}^2/Q^2)\to\log (m_{\tst_2}/m_{\tst_1})\to 1
\right)$, leading also to suppression.  So while large $|A_t|$ values
suppress both top squark contributions to $\Sigma_u^u$, at the same time
they also lift up the value of $m_h$, which is near maximal for large,
negative $A_t$.  Combining all effects, there exists the possibility
that the same mechanism responsible for boosting the value of $m_h$ into
accord with LHC measurements can also suppress EWFT, leading to a model
with electroweak naturalness.

\begin{figure}[tbp]
\postscript{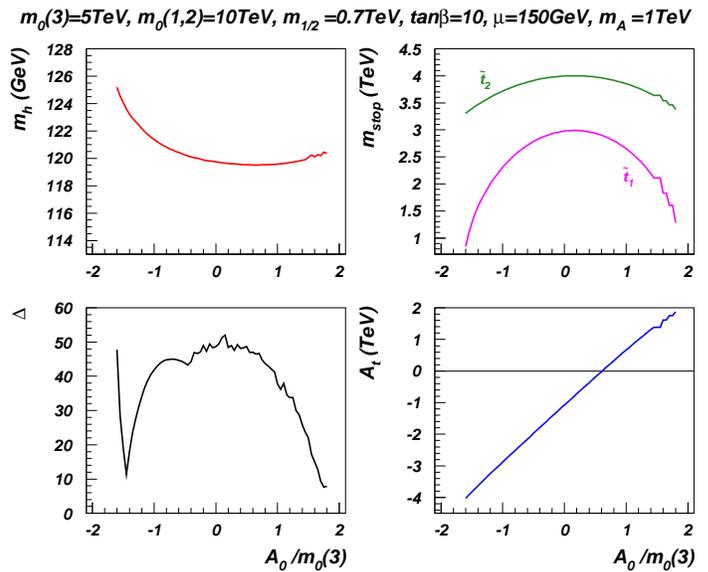}{1.1}
\caption{Plot of {\it a}). $m_h$, {\it b}). $m_{\tst_{1,2}}$, {\it c}). $\Delta$ and {\it d}). $A_t$ versus
variation in $A_0$ for a model with $m_0(1,2)=10$ TeV, $m_0(3)=5$ TeV, $m_{1/2}=700$ GeV, $\tan\beta =10$
and $\mu =150$ GeV and $m_A=1$ TeV. 
\label{fig:1}}
\end{figure}

To illustrate these ideas, we adopt a simple benchmark point from the 2-parameter non-universal Higgs mass
SUSY model NUHM2\cite{nuhm2}, but with split generations, where $m_0(3)<m_0(1,2)$. 
In Fig. \ref{fig:1}, we take $m_0(3)=5$ TeV, $m_0(1,2)=10$ TeV, $m_{1/2}=700$ GeV, $\tan\beta =10$ with
$\mu=150$ GeV, $m_A=1000$ GeV and $m_t=173.2$ GeV. We allow the GUT scale parameter $A_0$ to vary, and calculate the 
sparticle mass spectrum using Isajet 7.83\cite{isajet}, which includes the new EWFT measure. 
In frame {\it a})., we plot the
value of $m_h$ versus $A_0$. While for $A_0\sim 0$ the value of $m_h\sim 120$ GeV, as $A_0$ moves towards $-2m_0(3)$, 
the top squark radiative contributions to $m_h$ increase, pushing its value up to 125 GeV.
(There is an expected theory error of $\pm 2$ GeV in our RGE-improved effective potential calculation of
$m_h$, which includes leading 2-loop effects.\cite{hh})
At the same time, in frame {\it b})., we see the values of $m_{\tst_{1,2}}$ versus $A_0$. In this case, 
large values of $A_0$ suppress the soft terms $m_{Q_3}^2$ and $m_{U_3}^2$ via RGE running. But 
also large weak scale values of $A_t$ provide large mixing in the top squark mass matrix 
which suppresses $m_{\tst_1}$ and leads to an increased splitting between the two mass eigenstates 
which suppresses the top squark radiative corrections $\Sigma_u^u$. 
The EWFT measure $\Delta$ is shown in frame {\it c})., where we
see that while $\Delta\sim 50$ for $A_0=0$, when $A_0$ becomes large, then $\Delta$ drops to 10,
or $\Delta^{-1}= 10\%$ EWFT.
In frame {\it d})., we show the weak scale value of $A_t$ versus $A_0$ variation.
While the EWFT is quite low-- in the range expected for natural SUSY models-- we note that the top
squark masses remain above the TeV level, and in particular $m_{\tst_2}\sim 3.5$ TeV, in contrast to
previous natural SUSY expectations. 

The sparticle mass spectrum for this radiative NS benchmark point (RNS1)
is shown in Table \ref{tab:bm} for $A_0=-7300$ GeV. The heavier spectrum
of top and bottom squarks seem likely outside of any near-term LHC
reach, although in this case gluino\cite{bblt} and possibly heavy 
electroweak-ino\cite{wh} pair
production may be accessible to LHC14.  Dialing the $A_0$
parameter up to $-8$ TeV allows for $m_h=125.2$ GeV but increases EWFT
to $\Delta =29.5$, or 3.4\% fine-tuning. Alternatively, pushing $m_t$ up
to 174.4 GeV increases $m_h$ to $124.5$ GeV with 6.2\% fine-tuning;
increasing $\tan\beta$ to 20 increases $m_h$ to 124.6 GeV with 3.3\%
fine-tuning.  We show a second point RNS2 with $m_0(1,2)=m_0(3)=7.0$ TeV
and $\Delta =11.5$ with $m_h=125$ GeV; note the common sfermion mass parameter
at the high scale.  For comparison, we also show in
Table \ref{tab:bm} the NS2 benchmark from Ref. \cite{ns}; in this case,
a more conventional light spectra of top squarks is generated leading to
$m_h=121.1$ GeV, but the model-- with $\Delta=23.7$-- has higher EWFT
than RNS1 or RNS2.
%

%
\begin{table}\centering
\begin{tabular}{lccc}
\hline
parameter & RNS1 & RNS2 & NS2 \\
\hline
$m_0(1,2)$      & 10000 & 7025.0 & 19542.2  \\
$m_0(3)$      & 5000 & 7025.0 & 2430.6  \\
$m_{1/2}$  & 700 & 568.3 & 1549.3  \\
$A_0$      & -7300 & -11426.6 & 873.2  \\
$\tan\beta$& 10 & 8.55 & 22.1  \\
$\mu$      & 150 & 150 & 150  \\
$m_A$      & 1000 & 1000 & 1652.7  \\
\hline
$m_{\tg}$   & 1859.0 & 1562.8 & 3696.8   \\
$m_{\tu_L}$ & 10050.9 & 7020.9 & 19736.2  \\
$m_{\tu_R}$ & 10141.6 & 7256.2 & 19762.6  \\
$m_{\te_R}$ & 9909.9 & 6755.4 & 19537.2  \\
$m_{\tst_1}$& 1415.9 & 1843.4 & 572.0  \\
$m_{\tst_2}$& 3424.8 & 4921.4 & 715.4  \\
$m_{\tb_1}$ & 3450.1 & 4962.6 & 497.3  \\
$m_{\tb_2}$ & 4823.6 & 6914.9 & 1723.8  \\
$m_{\ttau_1}$ & 4737.5 & 6679.4 & 2084.7  \\
$m_{\ttau_2}$ & 5020.7 & 7116.9 & 2189.1  \\
$m_{\tnu_{\tau}}$ & 5000.1 & 7128.3 & 2061.8  \\
$m_{\tw_2}$ & 621.3  & 513.9 & 1341.2  \\
$m_{\tw_1}$ & 154.2  & 152.7 & 156.1  \\
$m_{\tz_4}$ & 631.2 & 525.2 & 1340.4  \\ 
$m_{\tz_3}$ & 323.3 & 268.8 & 698.8   \\ 
$m_{\tz_2}$ & 158.5 & 159.2 & 156.2  \\ 
$m_{\tz_1}$ & 140.0 & 135.4 & 149.2  \\ 
$m_h$       & 123.7 & 125.0 & 121.1  \\ 
\hline
$\Omega_{\tz_1}^{std}h^2$ & 0.009 & 0.01 & 0.006  \\
$BF(b\to s\gamma)\times 10^4$ & $3.3$  & 3.3 & $3.6$  \\
$BF(B_s\to \mu^+\mu^-)\times 10^9$ & $3.8$  & 3.8 & $4.0$  \\
$\sigma^{SI}(\tz_1 p)$ (pb) & $1.1\times 10^{-8}$  & $1.7\times 10^{-8}$ & $1.8\times 10^{-9}$ \\
$\Delta$ & 9.7 & 11.5 & 23.7 \\
\hline
\end{tabular}
\caption{Input parameters and masses in~GeV units
for two Radiative natural SUSY benchmark points and one NS point with $\mu =150$ GeV
and $m_t=173.2$ GeV.
}
\label{tab:bm}
\end{table}

\begin{figure}[tbp]
\postscript{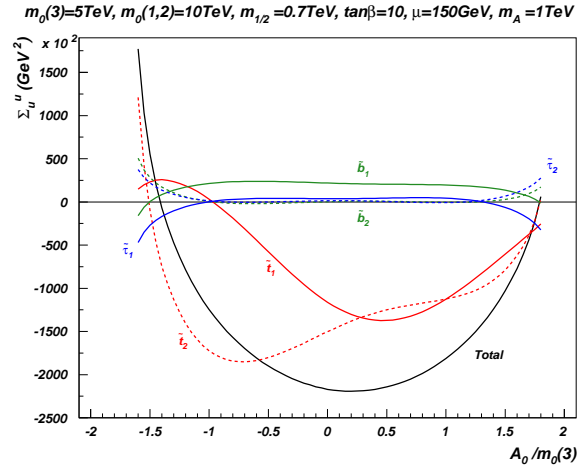}{0.9}
\caption{Plot of third generation contributions to $\Sigma_u^u$ versus $A_0$ for benchmark point RNS1 
where solid curves come form the lighter mass eigenstate and dashed curves from the heavier. 
The black solid curve is $\Sigma_u^u$ which has summed over 
all contributions. 
\label{fig:2}}
\end{figure}

To illustrate how low EWFT comes about even with rather heavy top
squarks, we show in Fig. \ref{fig:2} the various third generation
contributions to $\Sigma_u^u$, where the lighter mass eigenstates are
shown as solid curves, while heavier eigenstates are dashed. The sum of
all contributions to $\Sigma_u^u$ is shown by the black curve marked
total. From the figure we see that for $A_0\sim 0$, indeed both top
squark contributions to $\Sigma_u^u$ are large and negative, leading to
a large value of $\Sigma_u^u(total)$, which will require large
fine-tuning in Eq.~(\ref{eq:mssmmu}). As $A_0$ gets large negative, both
top squark contributions to $\Sigma_u^u$ are suppressed, and
$\Sigma_u^u(\tst_1 )$ even changes sign, leading to cancellations
amongst the various $\Sigma_u^u$ contributions.

\begin{figure}[tbp]
\postscript{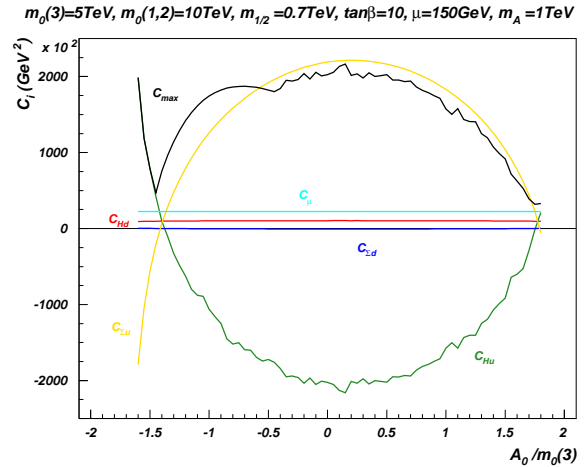}{0.9}
\caption{Various $C_i$ contributions to Eq.~(\ref{eq:mssmmu}) versus
$A_0$ for benchmark point RNS1.
\label{fig:3}}
\end{figure}

The overall effect on EWFT is exhibited in Fig. \ref{fig:3} where we plot several contributions $C_i$ to the RHS 
of Eq.~(\ref{eq:mssmmu}) versus $A_0$. Since $\mu$ is chosen close to $m_Z$, $C_\mu =(150\ {\rm GeV})^2$ 
is already quite small. 
The contribution $C_{\Sigma_u^u}\equiv -\Sigma_u^u\tan^2\beta/(\tan^2\beta -1)$ is large at $A_0\sim 0$, 
requiring a large value of $C_{H_u}\equiv -m_{H_u}^2\tan^2\beta/(\tan^2\beta -1)$ for cancellation to maintain
a small value of $\mu$. As $A_0$ becomes large negative, $C_{\Sigma_u^u}$ drops towards zero, so that only small
values of $C_{H_u}$ are needed to maintain $\mu=150$ GeV.

{\it Summary:} Models of Natural SUSY are attractive in that they enjoy
low levels of EWFT, which arise from a low value of $\mu$ and possibly a
sub-TeV spectrum of top squarks and $\tb_1$.  In the context of the
MSSM, such light top squarks are difficult to reconcile with the LHC
Higgs boson discovery which favors $m_h\sim 125$ GeV. Models with a
large negative trilinear soft-breaking parameter $A_t$ can maximize the
value of $m_h$ into the $125$ GeV range without recourse to adding
exotic matter into the theory. The large value of $A_t$ also suppresses
top squark contributions to the scalar potential minimization condition
leading to models with low EWFT and a light Higgs scalar consistent with
LHC measurements. 
(More details on the allowable parameter space of RNS will be presented 
in Ref. \cite{bbhmmt}.)
The large negative $A_t$ parameter can arise from
large negative $A_0$ at the GUT scale. In this case, large $A_0$ acts
via 1-loop renormalization group equations (RGEs) and large $m_0(1,2)$
acts through 2-loop RGEs\cite{am,gutimh} to squeeze multi-TeV third
generation masses down into the few TeV range, thus generating the
natural SUSY model radiatively. While RNS may be difficult to detect at
LHC unless gluinos, third generation squarks or the heavier
electroweak-inos are fortituously light, a linear $e^+e^-$ collider with
$\sqrt{s}\agt 2|\mu|$ would have enough energy to produce the hallmark
light higgsinos which are expected in this class of models.

{\it Acknowledgements:} 
HB would like to thank the Center for Theoretical Underground Physics
(CETUP) for hospitality while this work was completed.
This work was supported in part by the US Department of Energy, Office of High
Energy Physics.


%
\end{document}